\documentclass[conference]{IEEEtran}
\IEEEoverridecommandlockouts
\usepackage{cite}
\usepackage{amsmath,amssymb,amsfonts}
\usepackage{algorithmic}
\usepackage{graphicx}
\usepackage{textcomp}
\usepackage{xcolor}
\usepackage{hyperref}
\usepackage{tabularx}
\def\BibTeX{{\rm B\kern-.05em{\sc i\kern-.025em b}\kern-.08em
T\kern-.1667em\lower.7ex\hbox{E}\kern-.125emX}}
\begin{document}

\title{Controllable Singing Voice Synthesis \\ using Phoneme-Level Energy Sequence\\
}

\author{
\IEEEauthorblockN{Yerin Ryu, Inseop Shin, Chanwoo Kim}
\IEEEauthorblockA{
Department of Artificial Intelligence, Korea University, Seoul, Korea \\
\{yerinryu, vdlstjqn, chanwcom\}@korea.ac.kr}
}

\maketitle

\begin{abstract}
Controllable Singing Voice Synthesis (SVS) aims to generate expressive singing voices reflecting user intent. While recent SVS systems achieve high audio quality, most rely on probabilistic modeling, limiting precise control over attributes such as dynamics. We address this by focusing on dynamic control—temporal loudness variation essential for musical expressiveness—and explicitly condition the SVS model on energy sequences extracted from ground-truth spectrograms, reducing annotation costs and improving controllability. We also propose a phoneme-level energy sequence for user-friendly control. To the best of our knowledge, this is the first attempt enabling user-driven dynamics control in SVS. Experiments show our method achieves over 50\% reduction in mean absolute error of energy sequences for phoneme-level inputs compared to baseline and energy-predictor models, without compromising synthesis quality.

\end{abstract}


\section{Introduction}
Singing Voice Synthesis (SVS) is the task of generating singing voice waveforms from symbolic musical representations such as scores. While recent SVS systems have achieved substantial improvements in synthesis quality, most existing approaches generate expressive singing voices in a probabilistic manner conditioned only on the provided score. These existing systems lack the capability to provide users to precisely control or convey user intent and musical expression. Consequently, there has been increasing interest in developing controllable SVS frameworks that can explicitly  reflect user-specified expressive attributes. For example, SinTechSVS\cite{b1} demonstrated the synthesis of singing voices with eight distinct singing techniques. More recently, Prompt-Singer\cite{b2} proposed a system for controlling singing audio via natural language prompts, and TechSinger\cite{b3} further extended controllability by enabling user manipulation of various singing techniques through natural language instructions.

In this work, we focus on the controllable modeling of dynamics, a key expressive attribute in SVS. In the context of singing, As illustrated in Fig. 1, dynamics refers to the temporal variation of loudness, which is essential for conveying musical expressiveness\cite{b23}. The significance of dynamics has been widely acknowledged in the fields of music generation\cite{b5} and audio synthesis\cite{b6}. To the best of our knowledge, most existing SVS approaches\cite{b10, b11, b12, b13, b14, b15} have not explicitly incorporated dynamics as a controllable condition. \\

\begin{figure}[htbp]
  \centerline{
    \includegraphics[width=0.5\textwidth]{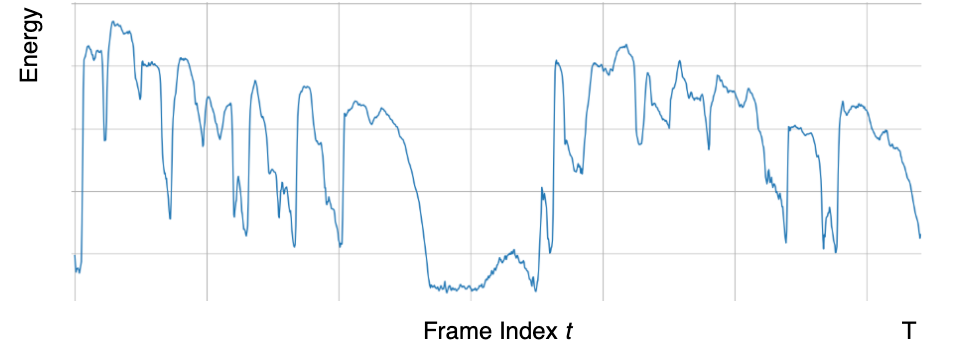}
  }
  \centerline{
    \text{(a) Energy Plot}
  }

  \vspace{1em}

  \centerline{
    \includegraphics[width=0.5\textwidth]{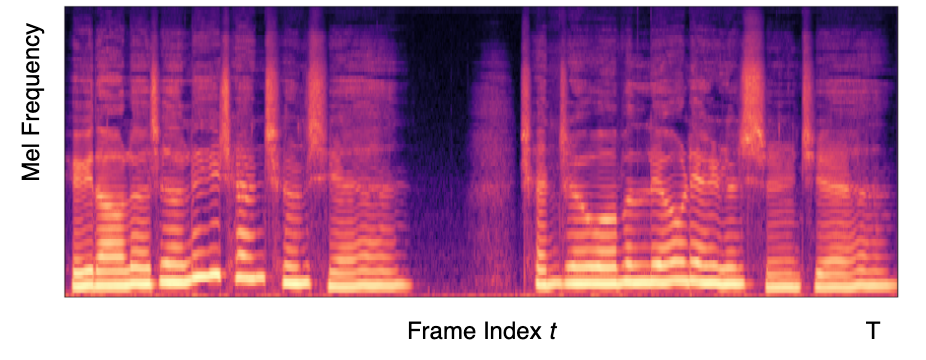}
  }
  \centerline{
    \text{(b) Mel-Spectrogram}
  }

  \caption{The energy plot in (a) represents the temporal variation of energy across time, calculated as the sum of mel-spectrogram amplitudes for each frame. When the energy is low, the corresponding regions in the mel-spectrogram (b) exhibit lower amplitude across frequency bins, appearing as darker. This indicates reduced loudness or silence in the audio signal during those time intervals.}
  \label{fig:1}
\end{figure}

According to \cite{b4}, dynamics is typically defined as a measure of the energy of an acoustic signal, computed based on the amplitude of the waveform. Since this energy can be extracted from the ground-truth waveform or spectrogram in most singing voice datasets, the need for manual annotation is significantly reduced. 

In this work, we present experimental results demonstrating that conditioning the model on an energy sequence enables effective control over the dynamics of the synthesized singing voice. Furthermore, we propose a method for caculating a phoneme-level energy sequence to facilitate user-friendly control, and validate its effectiveness for dynamic manipulation through experimental result. 

The main contributions of this paper are summarized as follows:

\begin{itemize}
    \item We show that dynamic control can be effectively achieved in a controllable SVS system by simply summing the energy sequence embedding with other input embeddings.
    \item We introduce a phoneme-level energy sequence representation to enable intuitive and user-friendly control.
    \item We empirically analyze the causal relationship between the input energy sequence and the dynamics of the synthesized singing voice.
\end{itemize}

Audio examples are available at \href{https://kongtory.github.io/DCSVS/}{\text{https://kongtory.github.io/DCSVS/}}, including the baseline system, ground-truth waveforms reconstructed with a vocoder, and our proposed SVS systems conditioned on energy at the frame and phoneme levels. \\

\section{Related Works}

\subsection{Singing Voice Synthesis}
SVS has seen significant improvements in quality enabled by advances in deep learning methodologies and large-scale singing datasets. Among the most influential studies in this field are DiffSinger\cite{b7} and VISinger\cite{b8}. 

DiffSinger employs a two-stage architecture featuring an encoder-decoder framework paired with a vocoder. The encoder generates a musical score embedding from input lyric, note, and duration sequences, while a diffusion-based decoder synthesizes mel-spectrograms conditioned on this embedding. A pre-trained vocoder converts the generated mel-spectrogram into waveform but is excluded from evaluation to focus solely on mel-spectrogram quality assessment.

VISinger, in contrast, is an end-to-end SVS system built upon Variational Inference with adversarial learning for end-to-end text-to-speech (VITS)\cite{b9}. In this architecture, the input sequences are encoded into a latent representation, from which the decoder directly generates the waveform. Additionally, an adversarial discriminator is applied to the latent space to enhance output quality. Building upon these two foundational studies as baselines, diffusion-based\cite{b10, b11, b12, b13} and VITS-based SVS models\cite{b14, b15} have continued to evolve, further advancing the quality of synthesized singing voices.

Inspired by DiffSinger, our work adopts a diffusion-based framework for SVS, utilizing a Denoising Diffusion Probabilistic Model (DDPM) \cite{b16} as the mel-spectrogram decoder. \\

\subsection{Controllable Singing Voice Synthesis}
The audio quality of SVS has advanced significantly, and recent research increasingly emphasizes not only sound quality but also expressiveness and user-centric control. Prompt-Singer\cite{b2} pioneered the first controllable SVS model, enabling users to adjust attributes such as singer gender, vocal range, and volume through natural language prompts. However, the range of controllable attributes remains limited, and challenges persist in controlling singing style.

SinTechSVS\cite{b1} and TechSinger\cite{b3} address these challenges by developing systems that manipulate singing style through vocal techniques. SinTechSVS introduced four timbre-related and four pitch-related technique sequences, successfully synthesizing singing voices with expert-annotated technique data. TechSinger further advanced this approach by enabling technique control via natural language input, utilizing a dataset annotated with singing techniques. As shown by theses studies, controlling singing style through vocal techniques represents a highly promising research direction.

While prior work has relied on annotated datasets, we propose a method for controlling singing style using the energy sequence—a technique attribute that does not require manual annotation. \\

\subsection{Energy Predictor}

The energy predictor is a module that implicitly incorporates energy information to achieve more natural speech synthesis. During training, this module is optimized by minimizing the Mean Squared Error (MSE) between the predicted energy and the ground-truth energy, thereby encouraging the model to reflect energy in the generated output. The first study to introduce an energy predictor was FastSpeech 2\cite{b17}, and since then, speech synthesis and SVS studies\cite{b18, b19, b20} have adopted this approach to achieve natural singing synthesis.

However, implicitly modeling musical dynamics using an energy predictor limits both the diversity and controllability of dynamics. Therefore, we propose a system that removes the energy predictor and instead uses energy sequence as an explicit input. \\

\section{Method}

\subsection{Model Architecture}\label{AA}

\begin{figure}[htbp]
\centerline{
\includegraphics[width=0.5\textwidth]{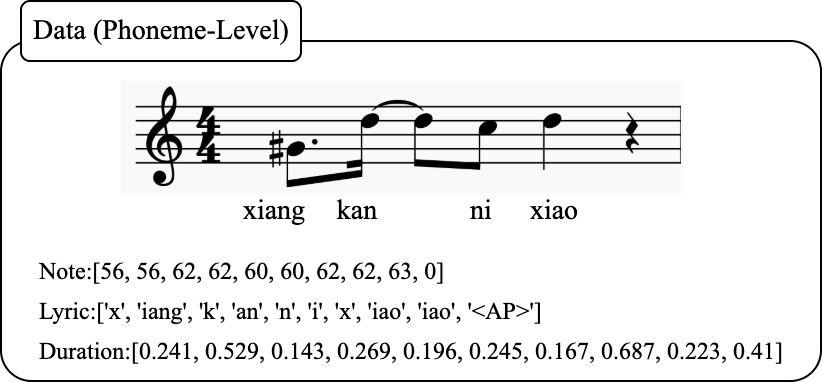}
}
\caption{Example of input sequences used in the SVS system. The first row shows the phoneme-level musical note sequence, the second row represents the corresponding phoneme-level lyric sequence, and the third row indicates the phoneme-level duration sequence. The duration sequence is annotated in seconds, and \textless{}AP\textgreater{} denotes a pause (rest) in the lyrics.}
\label{fig}
\end{figure}

\begin{figure*}[htbp]
  \centering
    \begin{minipage}[t]{0.6\textwidth}
        \centering
        \includegraphics[width=\textwidth]{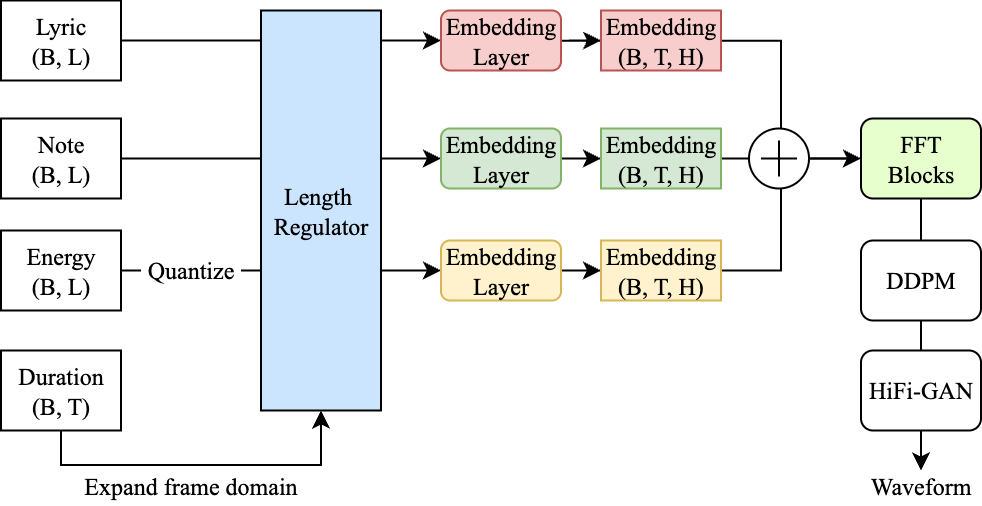}
        \vspace{0.5em}
        \text{(a) Overall Architecture}
    \end{minipage}
    \hfill
    \begin{minipage}[t]{0.35\textwidth}
        \centering
            \begin{minipage}[t]{0.48\textwidth}
                \centering
                \includegraphics[width=\textwidth]{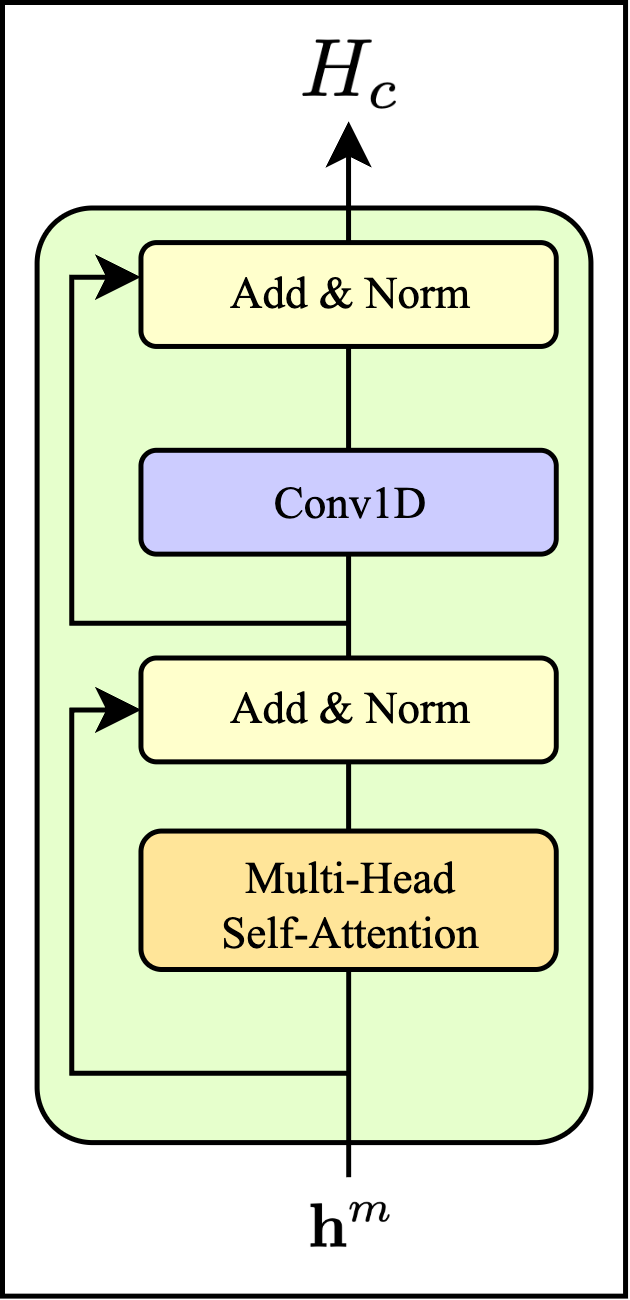}
                \vspace{0.5em}
                \text{(b) FFT Block}
            \end{minipage}
        \hfill
            \begin{minipage}[t]{0.48\textwidth}
            \centering
            \includegraphics[width=\textwidth]{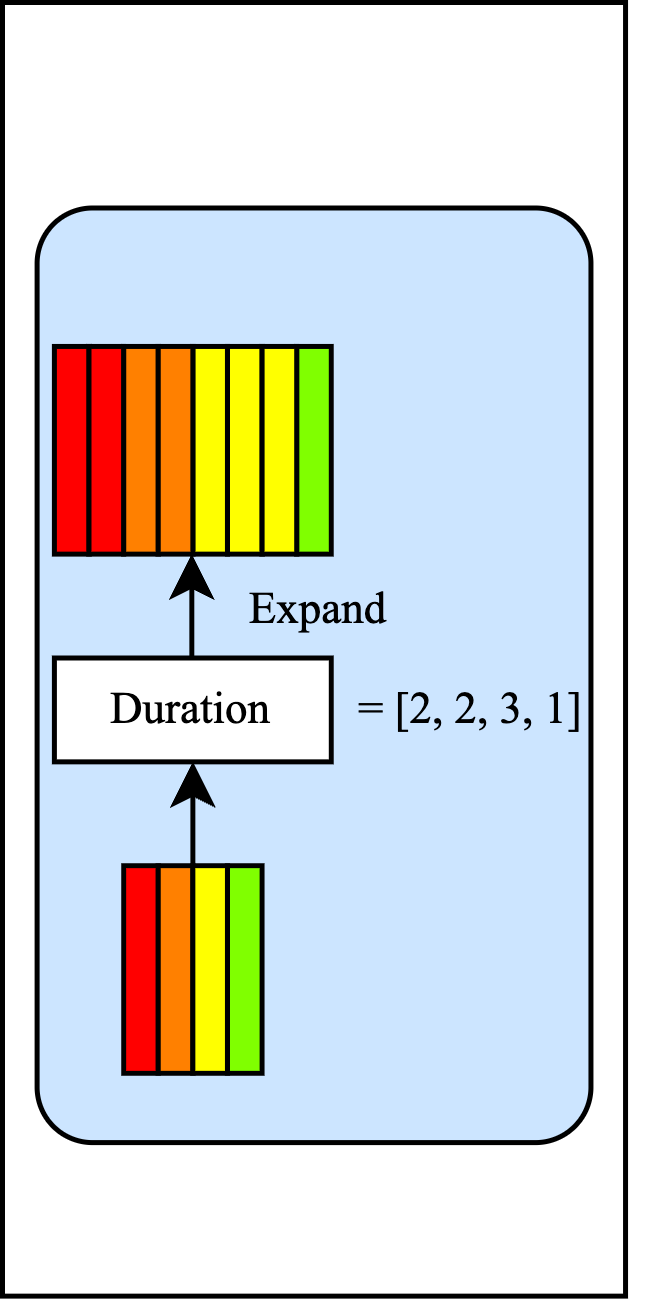}
            \vspace{0.5em}
            \text{(c) Length Regulator}
            \end{minipage}
    \end{minipage}

  \caption{Model architecture overview: (a) is overall architecture illustrating the input sequences of lyrics, notes, and phoneme-level energy, processed through the length regulator and FFT block, followed by the Denoising Diffusion Probabilistic Model (DDPM) mel-spectrogram decoder and vocoder. (b) is FFT block that sums the embeddings of all input sequences and applies a 2-layer 1D convolutional network to capture adjacent frame features. (c) is length regulator that expands phoneme-level sequences to frame-level sequences by repeating tokens according to duration annotations. Here, B denotes batch size, L denotes input sequence length, T denotes the expanded frame-level sequence length, and H denotes hidden size.}
  \label{fig:three_figs}
\end{figure*}

\begin{itemize}
    \item{Input Sequences}
\end{itemize}

In the SVS task, the input sequences consist of lyrics, musical notes, and duration. As shown in Fig. 2, all three sequences are annotated at the phoneme-level, and the length of each sequence is $L$. In this study, we additionally use a phoneme-level energy sequence of length $L$ as an input. The methodology is described in Subsection B. \\ \\

\begin{itemize}
\item{Length Regulator}
\end{itemize}

In speech synthesis, the duration predictor\cite{b17} is used to align the natural length of generated speech. Unlike speech synthesis, SVS datasets provide explicit duration annotations in seconds, enabling precise control over the target mel-spectrogram length. Given subtle individual variations in singing tempo, phoneme-level duration modeling better captures expressive nuances. 

In this study, as shown in Fig. 3(c), the duration annotations in seconds are converted to frame lengths, and each phoneme token is repeated according to its frame length via the length regulator\cite{b17}, producing a sequence of length $T$ for the target mel-spectrogram. The sequence is then converted into embedding vectors through the embedding layer. \\

\begin{itemize}
\item{FFT Block}
\end{itemize}

As shown in Fig. 3(b), the Feed-Forward Transformer (FFT) blocks \cite{b17} constructs the final hidden representation $H_c$, which will be used as the decoder input. A 2-layer 1D convolutional network is used to capture the features between adjacent frames. \\
\begin{itemize}
\item{Mel-Spectrogram Decoder}
\end{itemize}

The mel-spectrogram decoder generates the mel-spectrogram conditioned on $H_c$. In this study, we employ DDPM. \\

The model architecture can be expressed by the following equations:
\begin{equation}
\mathbf{h}^m = \mathbf{h}^l + \mathbf{h}^n + \mathbf{h}^e, \label{eq:arch1}
\end{equation}
\begin{equation}
H_c = \text{FFT}(\mathbf{h}^m), \label{eq:arch2}
\end{equation}
\begin{equation}
D_{out} = \text{DDPM}(H_c), \label{eq:arch3}
\end{equation}
\begin{equation}
\hat Y= \text{Vocoder}(D_{out}), \label{eq:arch4}
\end{equation}

where $\mathbf{h}^l$, $\mathbf{h}^n$, and $\mathbf{h}^e$ are the embedding vectors of the lyrics, note, and energy sequences, respectively, after passing through each embedding layer. By summing all the embedding sequences in this manner, $\mathbf{h}^m$ serves as a musical score embedding that comprehensively reflects the integrated input sequences. A pre-trained vocoder is used and excluded from the training stage. We fine-tuned HiFi-GAN\cite{b21} and used it for inference and evaluation. \\

\subsection{Input Energy Sequence}

In this section, we describe the phoneme-level energy sequence used as an input. \\

\subsubsection{Frame-Level Energy Eequence} We define the energy as the square root of the sum of energies across all channels rather than their direct sum to improve stability. We refer to this quantity simply as \textit{energy} throughout the paper. 

Energy can be extracted from the ground-truth mel-spectrogram without the need for human annotation. The length of the extracted energy sequence is equal to the number of frames $T$ in the mel-spectrogram, and we refer to this as frame-level energy:

\begin{equation}
\mathbf E[t] = \sqrt{\frac{1}{N} \sum_{n=0}^{N-1} \left( \exp(S_{\text{mel}}[t, n]) \right)^2 }.
\label{eq:energy}
\end{equation}

The log-mel spectrogram amplitude $S_{\text{mel}}[t, n]$ is derived from the magnitude-domain Short-Time Fourier Transform (STFT) as follows:
\begin{equation}
S_{\text{mel}}[t, n] = \log \left( \sum_{k=0}^{K-1} M[n, k] \cdot |S_{\text{stft}}[t, k]| \right),
\end{equation}
where $S_{\text{stft}}[t, k]$ denotes the complex STFT coefficient at frame $t$ and frequency bin $k$, $M[n, k]$ is the mel filterbank matrix, $|\cdot|$ denotes the magnitude operator, $t$ is the frame index, $n$ is the mel frequency bin index, $N$ is the total number of mel frequency bins, and $K$ is the number of STFT frequency bins. \\

\subsubsection{Phoneme-Level Energy Eequence}

We experimentally confirmed that when frame-level energy is added as an input embedding, the energy of the generated mel-spectrogram almost exactly follows the frame-level energy. However, using a frame-level energy sequence as input requires the user to specify and precisely align a very long sequence (on average, more than 1000 values for a waveform shorter than 10 seconds) with each token during inference, which is inconvenient for user control. Therefore, we chose to use the phoneme-level energy sequence as the input.

The phoneme-level energy sequence is defined as
\begin{equation}
\mathbf E = [e_1, e_2, ..., e_L]
\label{eq:phoneme_energy_seq},
\end{equation}
where $L$ is the sequence length, corresponding to the number of phonemes (identical to the lengths of the lyric and note sequences). 

The mean energy of the $i$-th phoneme, $e_i$, is computed as
\begin{equation}
e_i = \frac{1}{T_i} \sum_{t = t_{\text{start}_i}}^{t_{\text{end}_i}} \mathbf E[t],
\label{eq:phoneme_energy}
\end{equation}
where $T_i$ is the number of frames aligned to the $i$-th phoneme, $t_{\text{start}_i}$ and $t_{\text{end}_i}$ denote the start and end frame indices of the $i$-th phoneme, respectively, and $\mathbf E[t]$ is the frame-level energy at frame $t$. \\

\subsection{Objective Function}

We adopt the simplified objective function of the DDPM as our mel-spectrogram reconstruction loss. This objective minimizes the L1 loss between the noise predicted by the model at each timestep and the actual noise added to the data:

\begin{equation}
\mathcal{L} = \mathbb{E}_{\mathbf{x}_0, t, \epsilon}\left[ \left| \epsilon - \epsilon_\theta(\mathbf{x}_t, t, H_c) \right| \right],
\label{eq:simple}
\end{equation}

where $\mathbf{x}_0$ is the original data sample (ground-truth mel-spectrogram), $t$ is the diffusion timestep sampled uniformly from ${1, ..., T}$, $\epsilon$ is a noise vector sampled from a standard normal distribution $\mathcal{N}(0, I)$, $\mathbf{x}_t$ is the noisy version of $\mathbf{x}_0$ at timestep $t$, $\epsilon_\theta(\cdot)$ is the noise predicted by the neural network parameterized by $\theta$, and $H_c$ is the conditioning input.

The noisy data $\mathbf{x}_t$ is constructed as follows:
\begin{equation}
\mathbf{x}_t = \sqrt{\bar{\alpha}_t} \mathbf{x}_0 + \sqrt{1 - \bar{\alpha}_t} \epsilon,
\end{equation}
where $\bar{\alpha}_t = \prod_{s=1}^t (1 - \beta_s)$ is the cumulative product of the noise schedule, and $\beta_t$ is the variance schedule at timestep $t$ (we use linear scheduling).

The final objective function, including noise scheduling, is given by:
\begin{equation}
\mathcal{L} = \mathbb{E}_{\mathbf{x}_0, t, \epsilon}\left[ \left| \epsilon - \epsilon_\theta\left(\sqrt{\bar{\alpha}_t}\mathbf{x}_0 + \sqrt{1-\bar{\alpha}_t}\epsilon, t, H_c\right) \right| \right].
\label{eq:noise}
\end{equation}

\begin{figure*}[htbp]
  \centering
  \includegraphics[width=\textwidth]{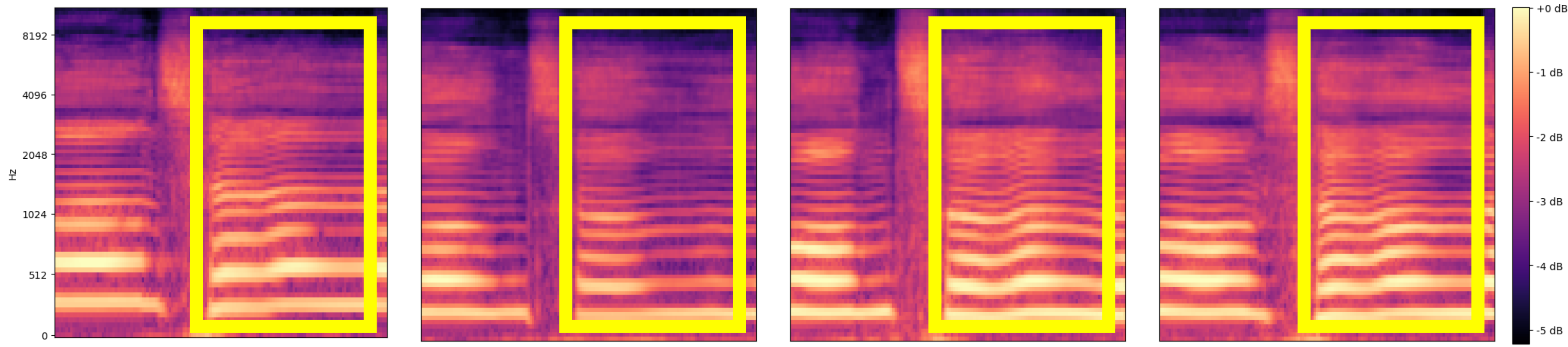} 

  \vspace{0.5em}

  \begin{tabularx}{\textwidth}{ 
    >{\centering\arraybackslash}X
    >{\centering\arraybackslash}X
    >{\centering\arraybackslash}X
    >{\centering\arraybackslash}X }
    \text{(a) Ground-Truth} & \text{(b) Baseline} & \text{(c) Phoneme-Level} & \text{(d) Frame-Level}
  \end{tabularx}

  \caption{Mel-spectrogram segments corresponding to the time region highlighted by the blue box in the Fig. 5.
(a) is ground-truth mel-spectrogram.
(b) is mel-spectrogram generated by the baseline model.
(c) is mel-spectrogram generated by the phoneme-level model.
(d) is mel-spectrogram generated by the frame-level model.
The yellow boxes indicate regions where the (b) exhibits noticeably lower energy (darker areas) compared to the (c) and (d). This visual difference demonstrates that incorporating the energy sequence as an input leads to increased energy in the synthesized output, confirming the effectiveness of explicit energy conditioning for dynamic control.}
  \label{fig:combined}
\end{figure*}

\begin{figure}[htbp]
  \centerline{\includegraphics[width=0.5\textwidth]{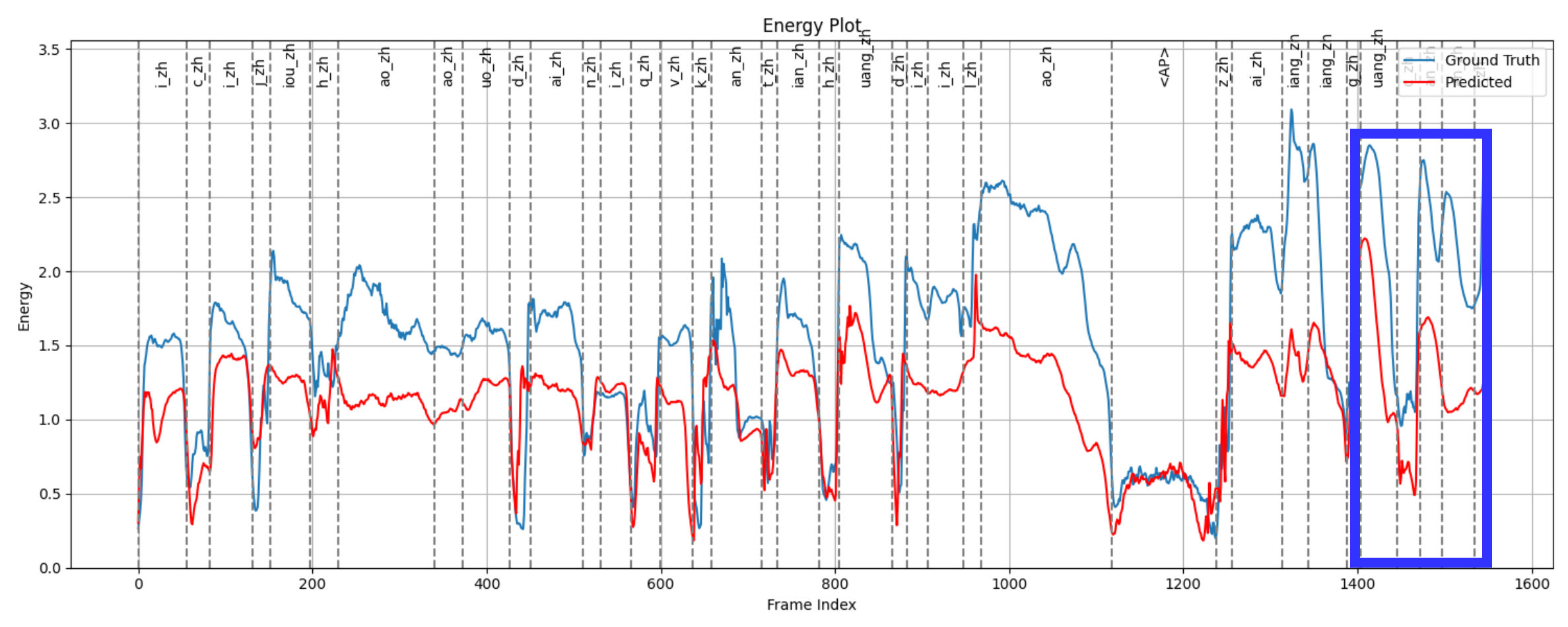}}
  \centerline{\text{(a) Baseline}}
  \vspace{1em}
  \centerline{\includegraphics[width=0.5\textwidth]{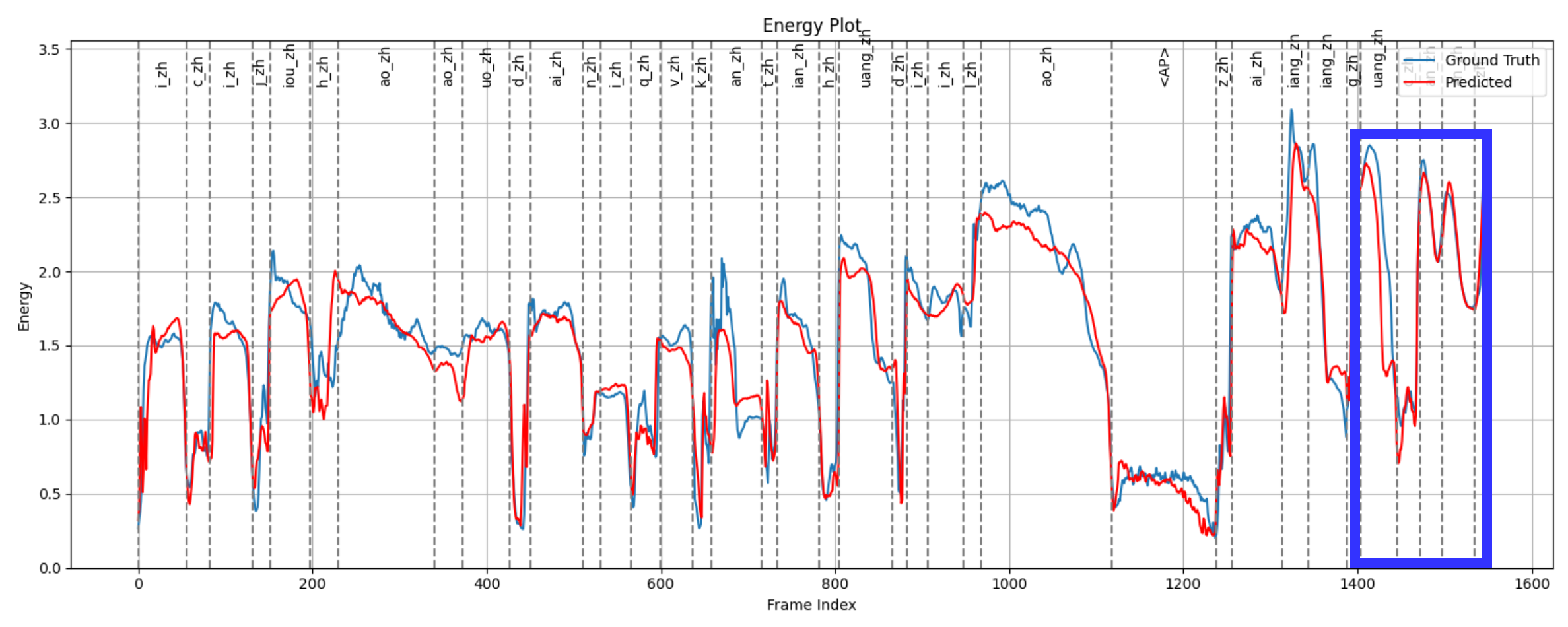}}
  \centerline{\text{(b) Phoneme-Level}}
  \vspace{1em}
  \centerline{\includegraphics[width=0.5\textwidth]{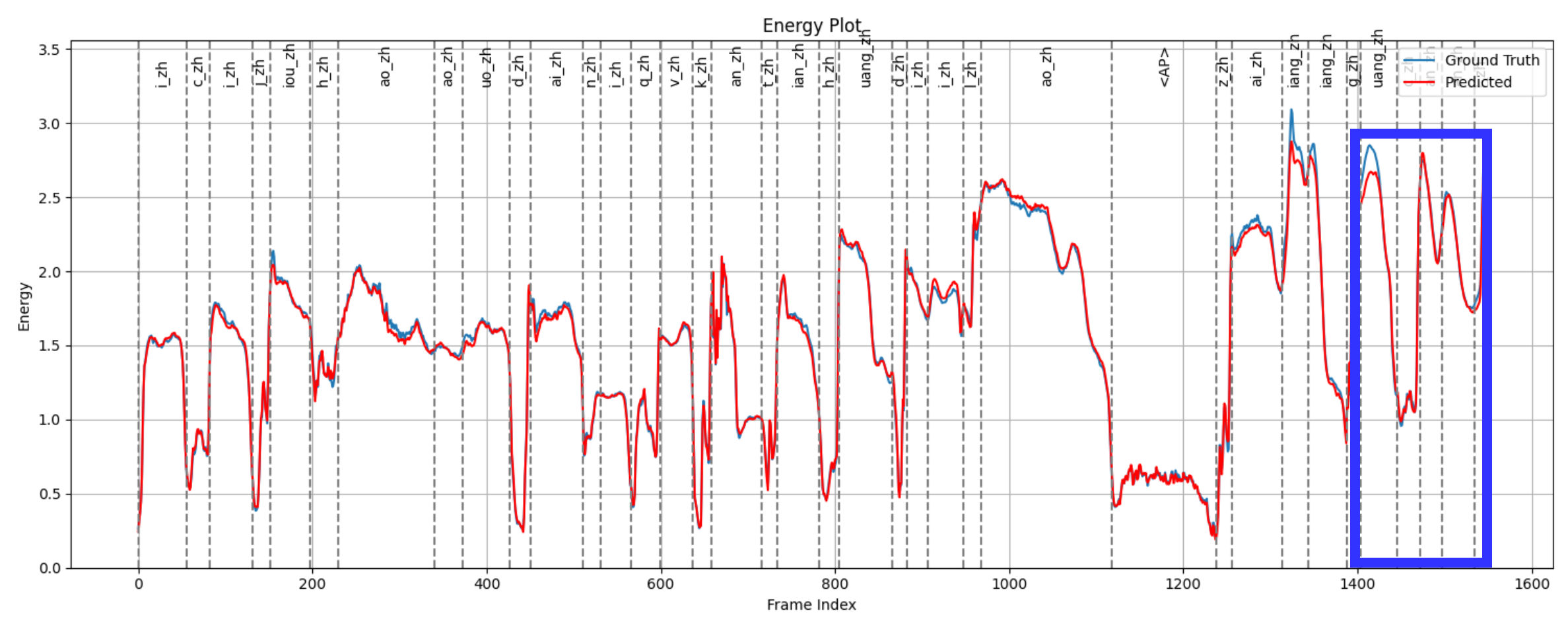}}
  \centerline{\text{(c) Frame-Level}}
  
  \caption{Energy plots comparing input energy sequences (blue) and generated mel-spectrogram energies (red) from: (a) baseline model, (b) phoneme-level model, and (c) frame-level model. The red curves in (b) and (c) closely follow the blue reference, demonstrating that explicit energy conditioning enables precise dynamic control in singing voice synthesis.}
  \label{fig:energy_plots}
\end{figure}

\section{EXPERIMENTAL SETUPS}
We used the GTSinger\cite{b22} dataset for our experiments. GTSinger is a high-quality singing voice dataset recorded with nine languages and six singing styles. Since our model is not designed for multilingual SVS, we trained it using only the Chinese subset, which contains approximately 16 hours of recordings from two speakers. The dataset was split into 7,082 training samples, 78 validation samples, and 57 test samples. Audio was sampled at 48 kHz, and for mel-spectrogram extraction, the hop size was set to 256, with both the window size and FFT size set to 1024. The number of mel-spectrogram bins was set to 80.

The model's sequence embedding dimension was set to 256, and relational positional encoding was applied before the FFT block. The embedding layer is a linear embedding layer for integer indices, which maps them into continuous vector representations. The mel-spectrogram decoder is a U-Net-based DDPM, configured with 100 forward steps, linear scheduling, and a maximum beta value of 0.06. All experiments were conducted on an RTX A5000 GPU, taking approximately 16 hours for 113,600 training steps. The vocoder used was a pre-trained HiFi-GAN, employ during inference and excluded from evaluation metrics.

For objective evaluation, we used mel-spectrogram-based metrics. F0 and energy were extracted from both the generated and ground-truth mel-spectrograms, and the Mean Absolute Error (MAE) was calculated for each. Objective evaluation metrics included F0 MAE, energy MAE, and Mel Cepstral Distortion (MCD). Subjective evaluation was conducted using the Mean Opinion Score (MOS). \\

\section{EXPERIMENTAL RESULTS}

The objective of our experiments is not to evaluate the audio quality itself, but rather to assess attribute controllability and the effect of energy sequence input on model performance. As a result, the MCD of the generated mel-spectrograms may be lower than State-Of-The-Art (SOTA) models; however, we confirmed that our method consistently outperforms the baseline in all experiments conducted in this study.

The baseline model generates mel-spectrograms using DDPM with only the lyric and note sequences as inputs. The frame-level model adds a frame-level energy sequence of length $T$ as an additional input to the baseline, while the phoneme-level model adds a phoneme-level energy sequence of length $L$ as an additional input to the baseline. \\

\subsection{Main Result}

\begin{table}[htbp]
\caption{Objective evaluation results for baseline, phoneme-level, and frame-level models. Both phoneme-level and frame-level energy conditioning significantly improve controllability and synthesis quality over the baseline.}
\begin{center}
\begin{tabular}{l|c|c|c|c}
\multicolumn{1}{c|}{Approach/Method} & Energy MAE $\downarrow$ & F0 MAE $\downarrow$ & MCD $\downarrow$ \\ \hline
Baseline       & 0.33 & 10.67 & 12.89 \\
Phoneme-Level  & \textbf{0.14} & \textbf{9.73} & \textbf{12.07} \\
Frame-Level    & 0.03 & 6.56 & 11.64 \\
\end{tabular}
\end{center}
\end{table}

\subsubsection{Energy MAE}
Energy MAE quantifies the MAE between the energy (amplitude) sequences extracted from generated mel-spectrogram and ground-truth mel-spectrogram. A lower value indicates superior fidelity in replicating energy patterns. As shown in the table 1, the baseline model has the highest energy MAE (0.33), while the phoneme-level model achieves a significant improvement (0.14). The frame-level model demonstrates the lowest error (0.03), establishing its dominance in precise energy control.

From the perspective of dynamic controllability, a higher energy MAE—indicating greater deviation from the exact replication of energy patterns—can reflect the fact that professional singers exhibit inherently diverse and inconsistent dynamics. Therefore, it suggests not only controllability of the energy sequence but also diversity in expressiveness.

The frame-level model’s energy curve, as visualized in Fig. 5(c), nearly perfectly mirrors the ground-truth energy profile. This visual evidence confirms its ability to not only replicate vocal dynamics with high precision but also reliably adhere to user-specified energy patterns.

Notably, while the phoneme-level model shows marginally higher errors than its frame-level counterpart, Fig. 5(b) reveals its robust capability to preserve the intended dynamic flow. This underscores the phoneme-level approach as a user-friendly control that maintains expressive quality without requiring fine-grained frame alignment.

Furthermore, the marginal improvements in F0 MAE and MCD suggest that incorporating the energy sequence as an additional input provides slight benefits to mel-spectrogram construction. \\

\subsubsection{MOS}

\begin{table}[htbp]
\caption{The Mean Opinion Score (MOS) with 95\% confidence intervals of test set. Both the phoneme-level and frame-level models achieved higher listening test scores than the baseline, demonstrating improved perceived audio quality.}
\begin{center}
\begin{tabular}{l|c}
\multicolumn{1}{c|}{}  & MOS $\uparrow$ \\ \hline
Ground-Truth + HiFi-GAN  & 4.02 $\pm$ 0.2 \\
Baseline + HiFi-GAN      & 3.43 $\pm$ 0.17 \\
Phoneme-Level + HiFi-GAN & 3.78 $\pm$ 0.19 \\
Frame-Level + HiFi-GAN   & 3.57 $\pm$ 0.18
\end{tabular}
\end{center}
\end{table}

The MOS results in Table 2 demonstrate that the ground-truth audio with HiFi-GAN vocoder achieves the highest score of 4.02 with a 95\% confidence interval of ±0.2. The baseline model with HiFi-GAN scores 3.43 ± 0.17, while the phoneme-level and frame-level models achieve MOS scores of 3.78 ± 0.19 and 3.57 ± 0.18, respectively. These results indicate that the phoneme-level model demonstrates promising performance in audio quality. \\

\subsection{Ablation Study}

\begin{table}[htbp]
\caption{Ablation study results comparing the baseline, baseline with energy predictor, and phoneme-level energy input models. Explicit energy input provides significantly better control over dynamics and improves synthesis quality compared to implicit energy prediction.}
\begin{center} \begin{tabular}{l|ccc} \multicolumn{1}{c|}{} & Energy MAE $\downarrow$ & \multicolumn{1}{l}{F0 MAE $\downarrow$} & \multicolumn{1}{l}{MCD $\downarrow$} \\ \hline
Baseline                    & \textbf{0.33}      & 10.67                      & 12.89                   \\
Baseline + Energy Predictor & \textbf{0.30}       & 9.43                       & 12.97                   \\
Phoneme-Level               & \textbf{0.14}       & 9.73                       & 12.07       
\end{tabular}
\end{center}
\end{table}

The Table 3. demonstrate that simply adding an energy predictor to the baseline model yields only marginal improvements in energy MAE (0.33 → 0.30). This suggests that the energy predictor, which models energy implicitly, has limited impact on controllability.

In contrast, the phoneme-level energy input model achieves a substantial reduction in energy MAE (0.14). These results clearly indicate that providing energy as an explicit input is significantly more effective for controlling the dynamics of the synthesized singing voice than relying on an implicit energy predictor. \\

\section{Conclusion}
In this work, we have demonstrated that explicit energy sequence conditioning enables effective and intuitive dynamic control in singing voice synthesis. By leveraging phoneme-level energy representations, our approach achieves significant improvements in controllability while maintaining synthesis quality. Although the baseline model employed in our experiments does not reflect SOTA performance, our results suggest that the proposed energy sequence input method can be readily integrated with more advanced SVS architectures to further enhance both controllability and expressive power. This paves the way for more natural, expressive, and user-controllable singing voice synthesis in future research.

Despite these results, several limitations remain. First, the MOS evaluation was conducted with only 10 participants, which may constrain the statistical reliability and generalizability of the subjective listening test results. Therefore, the focus of the MOS analysis should be on the fact that conditioning on the energy sequence does not lead to a degradation in the quality of the generated mel-spectrogram. As mentioned in Section I, the quality of the audio samples are available at \href{https://kongtory.github.io/DCSVS/}{\text{https://kongtory.github.io/DCSVS/}}. Secondly, our current framework primarily addresses dynamic control through energy modulation, leaving other expressive attributes—such as timbre, vibrato, and advanced singing techniques—relatively unexplored. As a final point, while phoneme-level energy representations offer a user-friendly interface, they may not fully capture the fine-grained temporal nuances inherent in natural singing performances. Future work will focus on combining energy with additional expressive features to achieve richer and more nuanced singing voice synthesis. \\

\section*{Acknowledgment}

This work was supported by the IITP(Institute of Information \& Coummunications Technology Planning \& Evaluation)- ITRC(Information Technology Research Center) grant funded by the Korea government(Ministry of Science and ICT)(IITP2025-RS-2024-00436857), IITP grant funded by the Korea government(MSIT) (No. RS-2019-II190079, Artificial Intelligence Graduate School Program(Korea University)), and IITP under the artificial intelligence star fellowship support program to nurture the best talents (IITP2025-RS-2025-02304828) grant funded by the Korea government(MSIT). \\

\end{document}